\documentclass{article}
\usepackage{spconf,amsmath,graphicx}

\usepackage{multirow}
\usepackage{caption}
\usepackage{amsmath}
\usepackage{tikz}
\usetikzlibrary{positioning,shapes}
\usepackage{booktabs}
\usepackage{graphicx}
\usepackage{subcaption}
\usepackage{times}
\usepackage{latexsym}
\usepackage{graphicx}
\usepackage[T1]{fontenc}
\usepackage[utf8]{inputenc}
\usepackage{microtype}
\usepackage{inconsolata}
\usepackage{amsfonts}
\usepackage{amssymb}
\usepackage{xcolor}
\usepackage{hyperref}
\usepackage{booktabs}
\usepackage{float}        

\usepackage{setspace}

\title{Learning Audio Concepts from Counterfactual Natural Language}

\name{
  Ali Vosoughi\textsuperscript{1$^{\star}$}\thanks{$^{\star}$The author performed the work during an internship at Bosch.},
  Luca Bondi\textsuperscript{2},
  Ho-Hsiang Wu\textsuperscript{2},
  Chenliang Xu\textsuperscript{1}\thanks{~~ This work was partially supported by the National Science Foundation (NSF) under Grant 1909912 and by the Defense Advance Research Projects Agency (DARPA) under HR00112220003. This paper does not necessarily reflect the position of the Government, and no official endorsement should be inferred.}\thanks{© ICASSP 2024 IEEE. Personal use of this material is permitted. Permission from IEEE must be obtained for all other uses, in any current or future media, including reprinting/republishing this material for advertising or promotional purposes, creating new collective works, for resale or redistribution to servers or lists, or reuse of any copyrighted component of this work in other works.}
}
\address{
  \textsuperscript{1}University of Rochester, Rochester, NY, USA \\
  \textsuperscript{2}Bosch Research, USA - Bosch Center for Artificial Intelligence
}

\begin{document}

\ninept
\maketitle
\begin{abstract}
Conventional audio classification relied on predefined classes, lacking the ability to learn from free-form text. Recent methods unlock learning joint audio-text embeddings from raw audio-text pairs describing audio in natural language. Despite recent advancements, there is little exploration of systematic methods to train models for recognizing sound events and sources in alternative scenarios, such as distinguishing \textit{fireworks} from \textit{gunshots} at outdoor events in similar situations. This study introduces causal reasoning and counterfactual analysis in the audio domain. We use counterfactual instances and include them in our model across different aspects. Our model considers acoustic characteristics and sound source information from human-annotated reference texts. To validate the effectiveness of our model, we conducted pre-training utilizing multiple audio captioning datasets. We then evaluate with several common downstream tasks, demonstrating the merits of the proposed method as one of the first works leveraging counterfactual information in audio domain. Specifically, the top-1 accuracy in open-ended language-based audio retrieval task increased by more than 43\%.
\end{abstract}
\begin{spacing}{0.92}
\vspace{-1em}
\section{Introduction}
\vspace{-0.5em}

Conventional audio processing in machine learning relies on predefined categories, limiting the ability of the models to understand audio nuances using descriptive text. These constraints limit open-ended and contrastive training for better audio-text alignment. New trends in the field improve classic models using audio-text learning from audio data and matching natural language descriptions that have become successful models in image-text tasks~\cite{radford2021learning}. Learning audio representations from pairs of audio and their textual descriptions facilitates the development of foundational models for audio tasks, enabling audio-text models to generalize beyond the confines of predefined classes, leveraging natural language descriptions~\cite{pengi}.

\begin{figure}[t]
    \centering
    \includegraphics[width=0.75\linewidth]{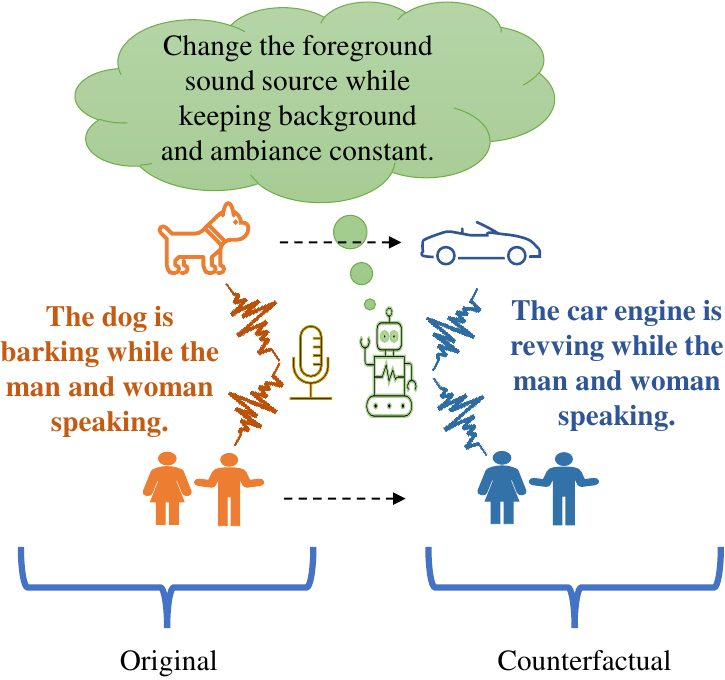}
    \caption{Counterfactual reasoning helps to distinguish various sound sources in an audio signal captured by a microphone. We identify isolated sound sources using \textit{GPT-3.5-Turbo}. Subsequently, we intervene to alter one or more of the sources of sounds to construct an imaginative linguistic representation in an alternative world that could have happened if there were other objects instead, thereby eliminating dependence on empirical audio data to reason objects driving the acoustic waves.}
    \label{fig:teaser}
    \vspace{-2em}
\end{figure}

Advancements in audio-text representation include AudioCLIP \cite{audioclip}, a tri-modal model, and Wav2CLIP \cite{wav2clip}, which extends CLIP \cite{radford2021learning} to audio. Subsequently, Elizalde \textit{et al.} proposed CLAP (Contrastive Language-Audio Pretraining) \cite{clap_paper}, a method that takes its inspiration from successful image-text models \cite{radford2021learning}. CLAP is able to train on audio and text directly without relying on image data for the learning process. However, human annotated audio captioning data is expensive and time-consuming to acquire. Recently, with large language models (LLMs) such as ChatGPT, an upgrade version of GPT-3 \cite{brown2020language} fine-tuned to follow human instructions \cite{ouyang2022training}, have become popular and been utilized to augment learning for various domains, both LAION-Audio-630K \cite{wu2023large} and WavCaps \cite{mei2023wavcaps} leverage the powerful textual re-writing and editing capabilities in order to acquire more audio-text pairs. Furthermore, methods such as Pengi \cite{pengi} and listen, think, and understand \cite{listen_think_understand} have built upon these foundational methodologies.

\begin{figure*}[t]
    \centering
    \begin{subfigure}{0.29\textwidth}
        \centering
        \includegraphics[height=1.9cm, width=\linewidth]{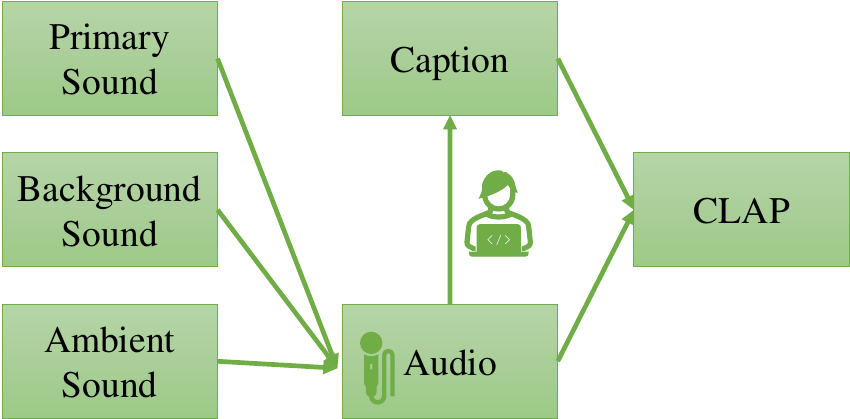}
        \caption{\small CLAP ~\cite{clap_paper}}
        \label{fig:sub1}
    \end{subfigure}
    \hfill
    \begin{subfigure}{0.65\textwidth}
        \centering
        \includegraphics[height=2.2cm, width=\linewidth]{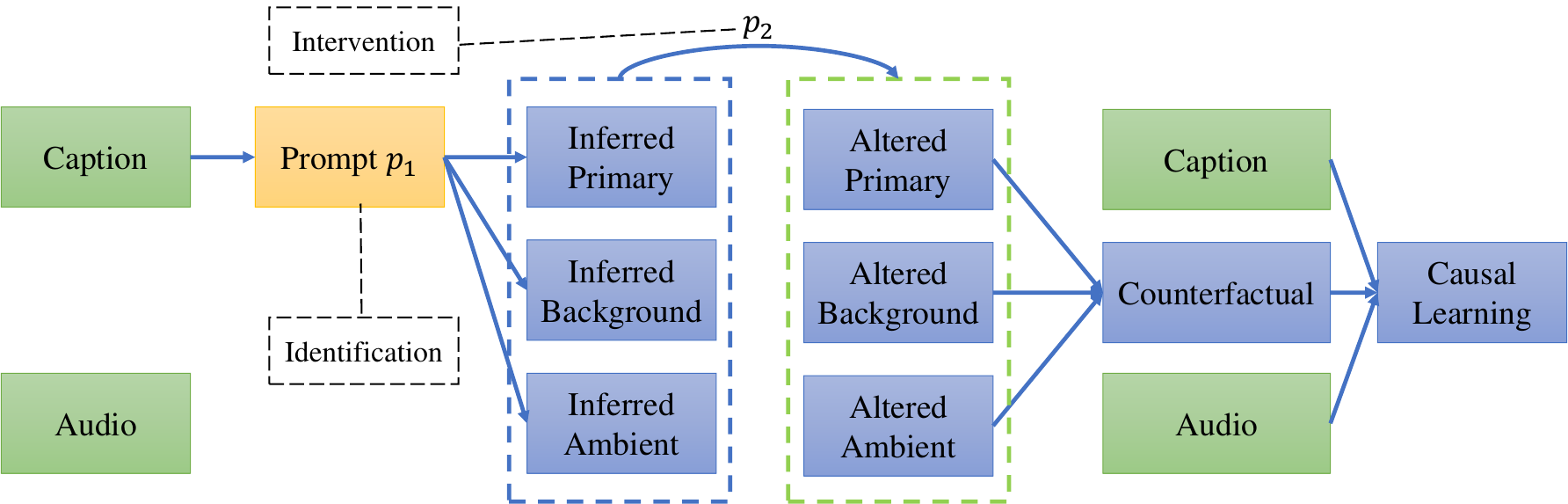}
        \caption{\small Our method}
        \label{fig:sub2}
    \end{subfigure}
    \hfill
    \caption{
   (a) The CLAP method~\cite{clap_paper} utilizes audio captions to train audio embeddings.
   (b) Our method leverages a prompt, represented by \( p = \{ p_1, p_2 \} \), using the GPT-3.5-Turbo model as LLM, that elicits counterfactual captions. This model guides interventions on the existing captions. The overarching goal is to pinpoint the sources of acoustic waves through human-narrated captions, shown as identification. After identifying these sources, controlling modification and intervention are performed by $p_2$ to incorporate the resulting counterfactual scenarios into a causal learning framework. This technique significantly improves the capability of audio-text models to distinguish subtle variations in text to align with sounds emanating from different objects. }
    \label{fig:2ndfigure}
    \vspace{-2em}
\end{figure*}

Distinguishing between sounds in similar conditions, such as the sound of a firework and a gunshot in the same concert event, requires controlled trials of both sounds and a learning algorithm to differentiate between them in the same context. However, existing audio-text datasets lack alternative scenarios for ethical and practical reasons. Counterfactual reasoning has been utilized to improve multimodal models involving vision modality \cite{zemni2023octet} and to aid grounding concepts within visual objects \cite{hendricks2018grounding}. The semantic differences between pairs of similar but slightly different audio clips have been discussed in \cite{kawaguchiaudio} and \cite{takeuchi2023audio}, addressing audio captioning. To the best of our knowledge, , our work is pioneering in utilizing the knowledge base and capabilities of LLMs to integrate counterfactual reasoning. This provides data and methods to enhance the learning of audio-text correspondence with counterfactual information. The proposed method utilizes counterfactual language and multimodal embeddings to improve textual discriminability in audio processing tasks, as shown in Figure~\ref{fig:teaser}. Subsequently, we develop a composite loss function that integrates the concept of triplet angular spaces and enforces factual audio-text consistency. We are inspired by recent developments in causality using LLMs~\cite{kiciman2023causal} that offer new possibilities, such as augmentation of meaningful counterfactuals in situations where audio learning data may not be available. The potential of the proposed method is to influence a wide array of applications, including automatic speech recognition, sound event detection, and audio-visual scene understanding. The core innovation of our work lies in being the first to integrate causal and counterfactual models into the analysis of audio data. Generated counterfactual sentences and prompts attempted will be available at \url{https://github.com/ali-vosoughi/counterfactual-audio}.

\vspace{-1em}
\section{Learning Audio from Counterfactual}
\vspace{-0.5em}
Causality describes the relationship between a cause and its subsequent effect. This study focuses on the causal relationships within audio samples that can be inferred from their corresponding text captions. Specifically, we address scenarios where data acquisition of counterfactual audio is impossible or costly by leveraging natural language as a substitute for imaginative data. Our method aligns with recent advancements in causality in natural language that aim to extend domains of causality to LLMs \cite{pearl2009causality, kiciman2023causal}. We first explore prior work of CLAP \cite{elizalde2023clap} that motivated us and then propose our method as the first work to extend counterfactuals to audio learning.

\noindent\textbf{Contrastive Language-Audio Pretraining (CLAP):}
Given a batch of \(N\) pairs \((x_i, y_i)\), \(x_i\) represents an audio sample and \(y_i\) is its corresponding text caption for \(i = 1, \ldots, N\). We define the audio and text encoders as \(E_a(i) = \phi_{\text{audio}}(x_i)\) and \(E_t(i) = \phi_{\text{text}}(y_i)\), respectively. These encoders transform each \(x_i\) and \(y_i\) into vectors in \(\mathbb{R}^{d}\), which are aggregated to form matrices \(E_a \in \mathbb{R}^{N \times d}\) and \(E_t \in \mathbb{R}^{N \times d}\). The similarity matrix \(C\) in the CLAP framework is given by  \eqref{eq:clap_similarity_loss} \cite{clap_paper}:

\vspace{-0.75em}
\begin{equation}
C = \tau \left( E_t \cdot E_a^\top \right)
\label{eq:clap_similarity_loss}
\end{equation}
The CLAP loss minimizes the discrepancy between audio and text representations, is denoted by \(\mathcal{L}\) and formulated as:
\vspace{-0.25em}
\begin{equation}
\mathcal{L} = 0.5 (\ell_{\text{text}}(C) + \ell_{\text{audio}}(C))
\end{equation}
Here, \(\tau\) serves as a scaling factor in  \eqref{eq:clap_similarity_loss}, modulating the effect of the similarity scores. While the original CLAP framework, shown in Figure~\ref{fig:sub1}, is valuable for audio-text pretraining, it does not allow causality expression. Addressing this gap, we introduce a counterfactual natural language strategy to infuse the CLAP framework with causality. This addition is beneficial as it circumvents the difficulties associated with data collection for counterfactual audio analysis.

\noindent\textbf{Causal Identification and Intervention:}
The counterfactual sentences in our model are generated through a prompt-based intervention on an observed caption \(y\), represented as \(y^\star = \text{do}(y|p)\). This prompt \(p\) is designed to fulfill three key aspects: it is factually grounded, identifies acoustic sources in captions to serve as identifying causes, and manipulates these sources to alter the caption, to serve as causal interventions. Identifiability means that all causal sources of acoustic waves can be obtained only from the language \cite{kiciman2023causal}. Examples of these transformations are depicted in Table \ref{tab:examples}.

\begin{table}[t]
\centering
\caption{Samples of original captions from the Clotho \cite{drossos2020clotho}, MACS \cite{macs}, and AudioCaps \cite{kim2019audiocaps} datasets and their counterfactual pairs.}
\label{tab:examples}
\begin{tabular}{@{}ll@{}}
\toprule
\multirow{2}{*}{\textbf{Dataset}} & \textbf{Original Caption} \\
& \textbf{Generated Counterfactual} \\
\midrule
\multirow{4}{*}{Clotho} & A gun is loaded, then loaded by hand some more \\
 & A piano is played, then played by hand some more. \\ \cmidrule{2-2}
 & A few gunshots are fired at the target shooting range \\
 & A few fireworks light up the night sky at shooting range. \\ 
 \midrule
\multirow{6}{*}{AudioCaps} & An adult male speaks and a crash occurs \\
 & An adult male speaks and a thunderstorm rumbles. \\ \cmidrule{2-2}
 & Large group of people clapping \\
 & Flock of birds chirping in unison. \\ \cmidrule{2-2}
 & Idling car, train blows horn and passes \\
 & Dogs barking, train blows horn and passes. \\ 
 \midrule
\multirow{6}{*}{MACS} & A crowd of people indoors talking \\
 & A group of cars honking on a busy street. \\ \cmidrule{2-2}
 & Adults and children are walking and talking \\
 & Cars and trucks are honking and zooming. \\ \cmidrule{2-2}
 & Adults talking and some footsteps coming across \\
 & Dogs barking and some footsteps coming across. \\
\bottomrule
\end{tabular}%
\vspace{-2em}
\end{table}

\noindent\textbf{Control Mechanisms for Counterfactual Language:}
We utilize prompts based on the Chain-of-Thought (CoT) method~\cite{wei2022chain} to align with objectives of causal identification and intervention and generating counterfactuals. Figure~\ref{fig:sub2} shows our prompt design's rationale, objectives, and context. We introduce a two-step prompting mechanism, denoted as \( p = \{ p_1, p_2 \} \), as shown in Fig. \ref{fig:sub2}. In this mechanism, \( p_1 \) anchors the discussion in factual elements, dissects acoustic objects, and plays the role of causal identifier. Concurrently, \( p_2 \) governs the generation of counterfactual statements by intervening in the identified causal acoustic sources. The decomposition of the audio captions to their acoustic sources and objects based on the physics of the acoustic waves using \( p_1 \) will ground the generated output of the LLM to physically possible scenarios. The counterfactuals can range from full negative examples, as found in hard negative sampling \cite{robinson2020contrastive}, to minor, physically plausible counterfactual scenarios. The steering of counterfactuals will influence the learning of the audio-text embeddings. For instance, in the example of "children and adult voices with footsteps and birds singing in the background," \( p_2 \) can be used to change the primary sound, inferred from the caption, "children and adult voices," background sound, "with footsteps," or ambient sound, "birds singing in the background." In this paper, we use a combination of these assumptions without restricting the assumption for the sake of generalizability. 

\noindent\textbf{Loss Functions to Incorporate Counterfactuals:}
The \textit{angle loss} aims to minimize the angular difference between factual and counterfactual captions. It is defined as:
\vspace{-0.75em}
\begin{align}
L_{\text{angle}} = \frac{1}{N}\sum_{i=1}^{N} \max\Big(0, \cos(\phi_{\text{audio}}(x_{i}), \phi_{\text{text}}(y^\star_{i})) \\
- \cos(\phi_{\text{audio}}(x_{i}), \phi_{\text{text}}(y_{i})) + \mu\Big)
\end{align}
Here, \(\mu\) represents the angular margin. To encourage factual consistency between audio samples and their corresponding captions, we define the \textit{factual consistency loss} as follows:
\vspace{-0.75em}
\begin{align}
L_{\text{factual\_consistency}} = \frac{1}{N}\sum_{i=1}^{N} \lVert \phi_{\text{audio}}(x_i) - \phi_{\text{text}}(y_i) \rVert^2_2 
\end{align}
The \textit{total loss}, combining both the factual consistency loss and angle loss, is expressed in  \eqref{eq:total_loss}:
\vspace{-0.75em}
\begin{equation}
L_{\text{total}} = w_1 \cdot L_{\text{angle}} + w_2 \cdot L_{\text{factual\_consistency}}
\label{eq:total_loss}
\end{equation}
Choices of hyperparameters in  \eqref{eq:total_loss} are pragmatic, serving as a trade-off to best exploit the capabilities of counterfactuals while ensuring factual consistency.

\vspace{-1em}
\section{Experimental design}
\vspace{-0.5em}

\subsection{Encoders} 

\noindent\textbf{Audio encoder:} We use PANNs \cite{kong2020panns} encoder, specifically ResNet-38 has been used with pretrained weights loaded with adapter layers to fine-tune model and align the embeddings.

\noindent\textbf{Text encoders:} We use the same CLIP text encoder modules \cite{radford2021learning} provided from HuggingFace\footnote{\url{https://huggingface.co/docs/transformers/model_doc/clip}} \cite{wolf2020transformers} for encoding captions and counterfactuals. The weights of the encoders were frozen in all stages. Therefore, we excluded the effect that may arise form the encoding performance of the language encoders.

We employed logarithmic Mel spectrograms of audio sampled at 32kHz. The hop size is set to 320 frames, the window size to 1024 frames, and we utilized 64 Mel bins spanning the frequency range of 50-14000 Hz. Audio clips were randomly truncated to contiguous 10-second segments for training purposes, with zero padding applied for shorter clips. The captions remained unaltered. During training, batches containing pairs of audio and text are randomly selected.

\vspace{-1em}
\subsection{Data}
\textbf{Training datasets:} Total of 44,292 from AudioCaps \cite{kim2019audiocaps}, 29,646 pairs from Clotho \cite{drossos2020clotho} (each audio has five captions, so we created five pairs per clip), and 17,276 pairs from MACS \cite{macs} have been used during pretraining. The reason that we picked these three datasets are that these are purely annotated by humans.

\noindent\textbf{Test datasets:} We use the \textit{Clotho} dataset \cite{drossos2020clotho} for evaluating the model's performance in language-based audio retrieval task. We generated counterfactual captions using the GPT-3.5-Turbo, similar to the training sets. For evaluating the model's performance as zero-shot classification in conventional problems with limited classes, we use the \textit{Environmental Sound Classification 50 (ESC-50)} \cite{piczak2015esc} with 50 predefined categories related to audio classes, and \textit{UrbanSound8K (US8K)} \cite{salamon2014dataset} with ten classes as {air conditioner, car horn, children playing, dog bark, drilling, engine idling, gunshot, jackhammer, siren, street music}.

\vspace{-1em}
\subsection{Baseline}
We adopt the approach from CLAP \cite{clap_paper}, and train our version with the same datasets we used in generating counterfactuals, including only AudioCaps, Clotho, and MACS.

\vspace{-1em}
\section{Results and Discussions}
\vspace{-0.5em}

\subsection{Evaluation on Downstream Tasks}

\textbf{Results on Clotho:} We use Clotho \cite{drossos2020clotho} to test the performance of our method on language-based audio retrieval task. As listed in Table \ref{tab:algorithm_comparison}, our method's performance on text-based retrieval tasks yields a 43\% improvement in top-1 accuracy, reinforcing its superior precision. Notably, the performance has slight improvement for top-10 retrieval tasks. We used the cosine similarity of text and audio embeddings to measure the performance. Therefore, the text encoder plays a crucial role in capturing small spelling nuances that have significant effects, and the limitations of the text encoder in capturing nuances will impact the model's performance. For instance, the cosine similarity of embeddings of "This is the sound of a dog" vs. "This is the sound of a cat" using BERTscore text encoder \cite{zhang2019bertscore} is 0.99, while these two sentences are carrying two significantly distinct animals. While this study excludes the effects of different text encoders to focus on the merits of counterfactual audio analysis, a comprehensive token dictionary and text encoder can improve sentence distinction.

\begin{table}[htbp]
\centering
\caption{\small Performance (out of 1) on the Clotho evaluation set for text-to-audio retrieval is listed.}
\vspace{-1em}
\begin{tabular}{@{}lcc@{}}
\toprule
\textbf{Method} & \textbf{Top-1} & \textbf{Top-10} \\
\midrule
CLAP & 0.088 & 0.395 \\
Our method & \textbf{0.126} & \textbf{0.423} \\
\bottomrule
\end{tabular}%

\label{tab:algorithm_comparison}
\end{table}

\noindent\textbf{Results on ESC-50 and US8K:} We evaluated the zero-shot classification performance of our proposed model on two benchmark datasets, ESC-50 and US8K, and summarized in Table \ref{tab:zero_shot_comparison}. As the table shows, our model performs commendably on the ESC-50 dataset, which features many classes. This performance is slightly better than that of the CLAP method. Conversely, the performance lags on the US8K dataset when compared to CLAP. One reason might be that the number of classes in US8K is much lower, in contrast to what our model learned during the training. As the class labels of the US8K lack sufficient textual detail about the data, the classification becomes harder as all labels get relatively similar and high similarity scores. For instance, the BERTscore distance \cite{zhang2019bertscore} between class label "siren" and all other class labels is 0.844 $\pm$ 0.020, and the trend consistently repeats for all class labels of the dataset. Contrastively, for the ESC-50 dataset, the value of similarity of the class label "siren" to all other classes is 0.821 $\pm$ 0.018.  
Further analysis is necessary to ascertain the reasons behind the reduced performance on the US8K dataset and to refine our model for enhanced overall accuracy.

\begin{table}[htbp]
\centering
\caption{Comparison of zero-shot performance between our proposed method and existing methods on ESC-50 \cite{piczak2015esc} and US8K \cite{salamon2014dataset}.}
\vspace{-1em}
\begin{tabular}{@{}lcc@{}}
\toprule
\textbf{Method} & \textbf{ESC-50} & \textbf{US8K} \\
\midrule
Wav2CLIP \cite{wu2022wav2clip} & 0.414 & {0.404} \\
AudioClip \cite{guzhov2022audioclip} & 0.694 & {0.653} \\
CLAP & 0.729 & \textbf{0.798} \\
Our method & \textbf{0.744} & 0.475 \\
\bottomrule
\end{tabular}%

\label{tab:zero_shot_comparison}
\vspace{-1.5em}
\end{table}

\subsection{Ablation Studies}
\begin{figure}[t]
  \centering

  \begin{subfigure}[b]{0.225\textwidth}
    \includegraphics[width=\textwidth]{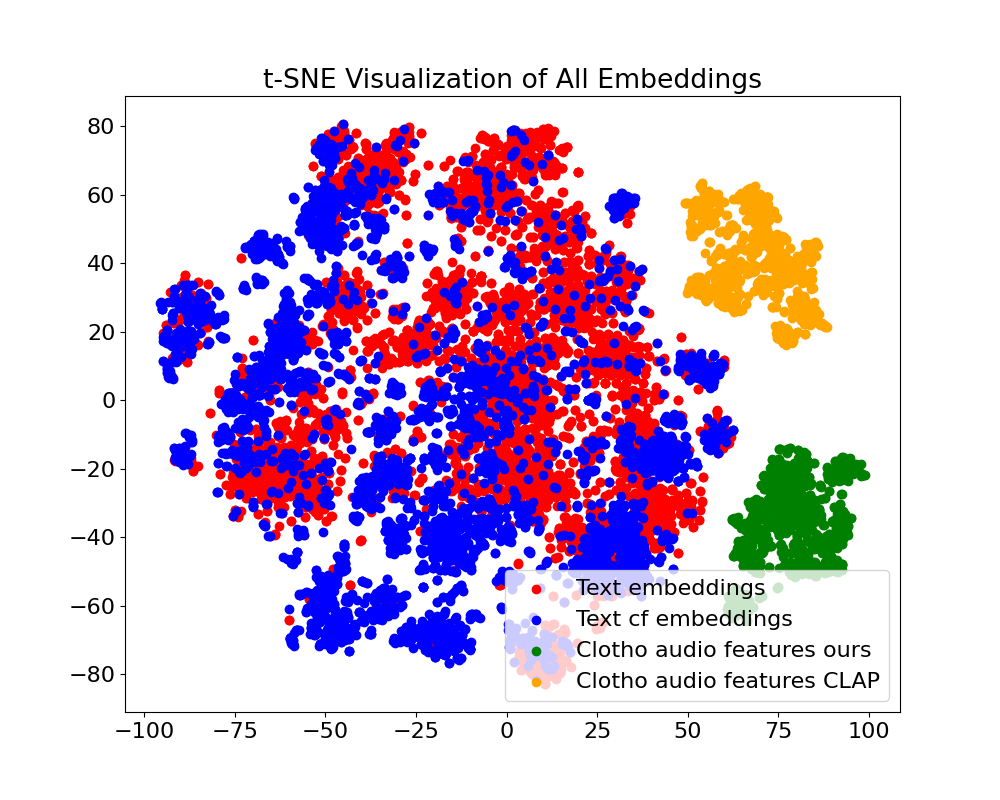}
    \vspace{-0.3in}
    \caption{\footnotesize $w_1$: 0, $w_2$: 0}
  \end{subfigure}
  \hfill
  \begin{subfigure}[b]{0.225\textwidth}
    \includegraphics[width=\textwidth]{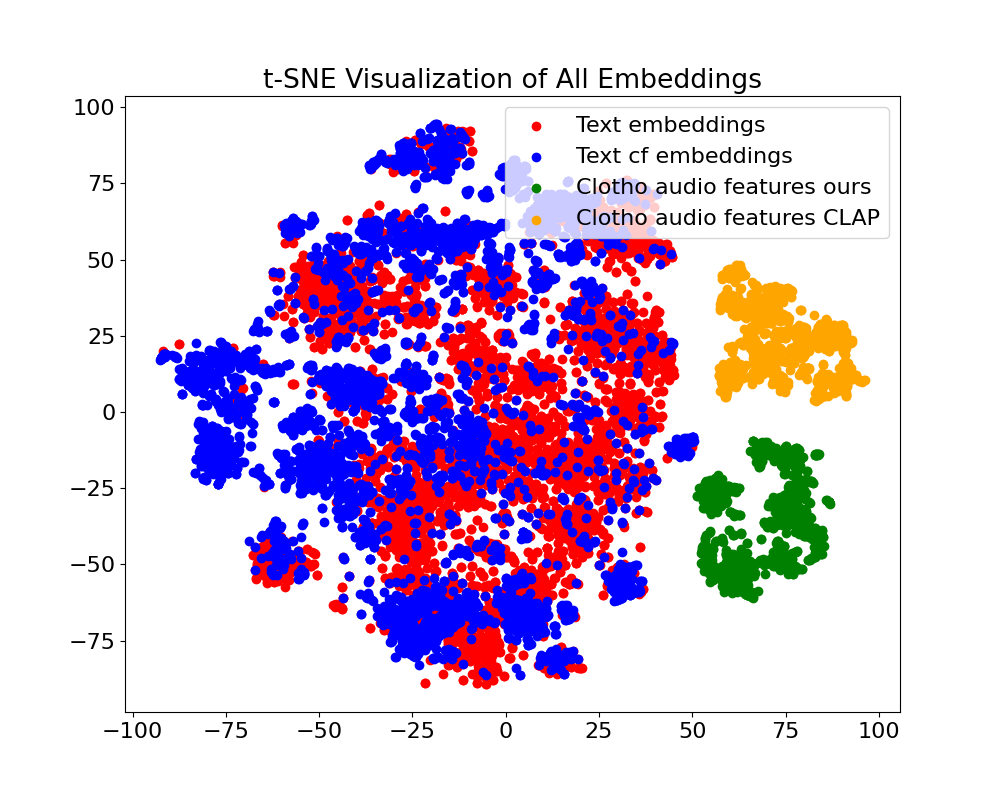}
    \vspace{-0.3in}
    \caption{\footnotesize $w_1$: 1, $w_2$: 0}
  \end{subfigure}

  \begin{subfigure}[b]{0.225\textwidth}
    \includegraphics[width=\textwidth]{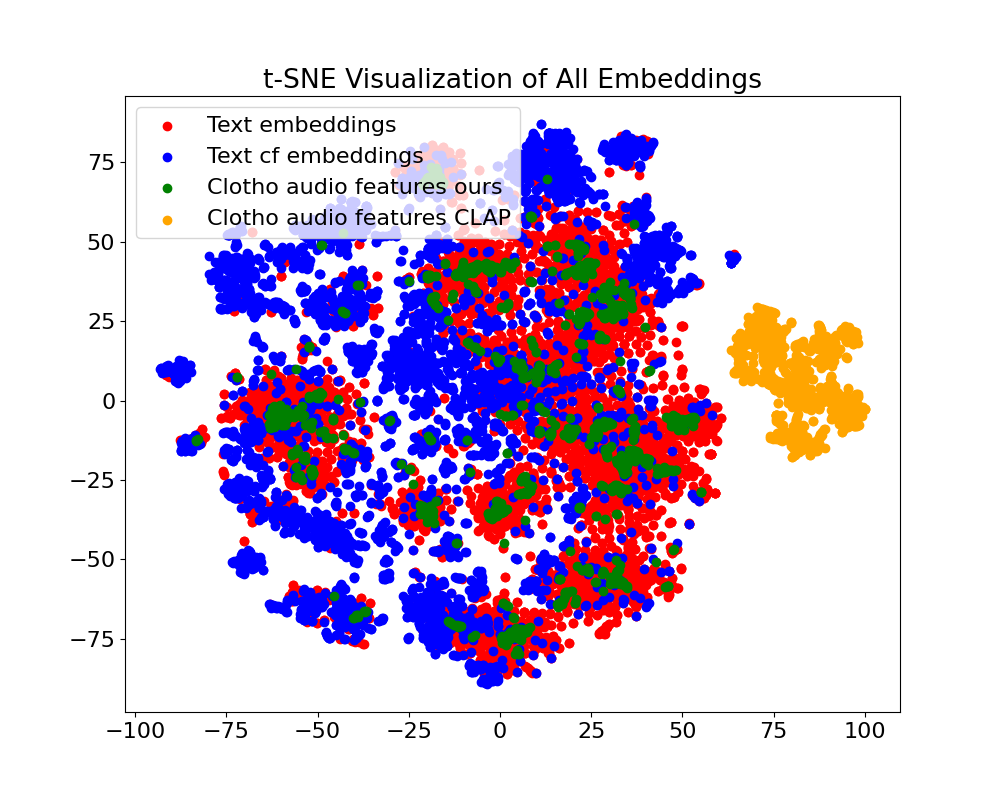}
    \vspace{-0.3in}
    \caption{\footnotesize $w_1$: 0, $w_2$: 100}
  \end{subfigure}
  \hfill
  \begin{subfigure}[b]{0.225\textwidth}
    \includegraphics[width=\textwidth]{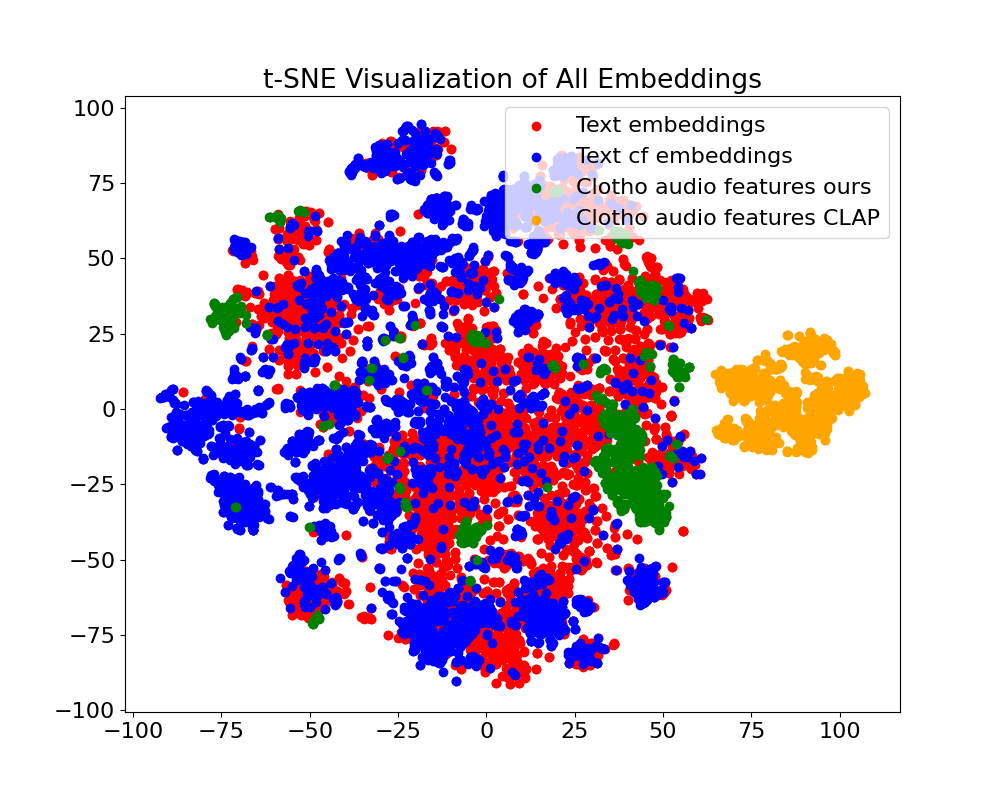}
    \vspace{-0.3in}
    \caption{\footnotesize $w_1$: 1, $w_2$: 100}
  \end{subfigure}
  \caption{t-SNE visualization of audio embeddings with ours and CLAP compared to original and counterfactual caption embeddings under different parameter configurations. Visualization keys: factual text (red dots), counterfactuals (blue), our audio embeddings (green), and CLAP audio embeddings (orange). As loss terms are incrementally introduced, our audio embeddings consistently align more closely with factual data and distance from counterfactuals. Our audio embeddings get closer to facts and distance from the counterfactuals for various combinations of angle loss and factual consistency loss terms.}
  \label{fig:tsne}
\vspace{-2em}
\end{figure}

Figure \ref{fig:tsne} shows the t-SNE visualizations of the embeddings for various settings of the coefficients. Evidently, starting from a random guess, the embeddings evolve with each subsequent addition of the loss terms. The plots in Fig. \ref{fig:tsne} show that when all loss functions are set to zero, the audio encoders are based on the PANN embeddings in ours, and text encoders which are frozen and are showing the fixed space of captions and counterfactuals. One particular observation is that audio embeddings of the CLAP are closer to the fact (orange dots) than counterfactuals (blue dots). This observation, by itself, shows that CLAP is favorably learning to stay closer to facts. One reason is that our generated counterfactuals combine various types of counterfactuals, and some of the counterfactuals are similar to negative, so CLAP can successfully work on them. By incorporating counterfactuals via angle loss, we observe that the audio embeddings of our method get distant from the counterfactuals, staying closer to the facts. Alternatively, by only including factual loss, audio embeddings tend to align fully with factual embeddings, staying closer to the facts. Noteworthy is the trade-off that adding the factual consistency loss reduces the distance to the counterfactuals; however, the distance to the facts becomes much less, and we observe this as a trade-off. Having both factual consistency loss and angle loss ensures that the counterfactuals are distant enough while facts are kept closer to the audio embeddings.

\begin{table}[ht]
\centering
\resizebox{0.4\textwidth}{!}{%
\begin{tabular}{@{}cccccc@{}}
\toprule
$w_1$ & $w_2$ & Top-1 & Top-10 & $sim(x,y)>sim(x,y^\star)$ \\
\midrule
0 & 0 & 0.0002 & 0.01 & $517.80 \pm 237.36$ \\
1 & 0 & 0.0819 & 0.3365 & $1000.80 \pm 5.76$ \\
0 & 100 & 0.1328 & 0.4379 & $861.80 \pm 293.76$ \\
1 & 100 & 0.1102 & 0.3782 & $967.80 \pm 43.36$ \\
\bottomrule
\end{tabular}%
}
\caption{Ablations on different combinations of $w_1$ and $w_2$. The last column represents the effect of each loss on the variation in audio-text similarity versus audio-counterfactual by changing the loss terms.}
\label{tab:ablation}
\vspace{-1em}
\end{table}

In order to quantify the closeness of audio embeddings to original text and counterfactual text, we use the cosine similarity of audio-text and audio-counterfactual embeddings. Based on these similarities, we count the number of times over 1044 samples of the Clotho evaluation set that the audio embeddings are closer to facts than counterfactuals. As expected in the first row, the random guess is 517, roughly about half of the 1044 samples in the test set. In Table \ref{tab:ablation}, adding angle loss \(w_1\) and using counterfactuals improve our audio embeddings. Another pattern is that having factual consistency loss improves accuracy; however, the closeness of audio-text, as compared to counterfactuals, is not as good as when we add angle loss. Therefore, by observing patterns in Fig. \ref{tab:ablation} and Table \ref{tab:ablation}, there is a trade-off between facts and counterfactuals. By getting closer to facts through factual consistency loss, the accuracy of Clotho retrieval increases, while the model does not learn to distinguish well between the facts and counterfactuals. In contrast, adding angle loss incorporates counterfactuals in training, helping to distinguish between descriptions that might be directly (as facts/captions) or indirectly (as counterfactual captions) related to the same audio. This trade-off is conformal with intuition. Imagine a judge evaluating two cases. One case presents strong evidence (facts), and the other presents a scenario that might have happened but did not (counterfactual). If a judge focuses solely on ensuring the judgments align perfectly with the facts, then might struggle to distinguish between the two cases effectively. However, considering both the facts and the counterfactual scenario, the judge becomes better at discerning their nuances. This balance between staying close to facts and acknowledging counterfactuals is what our model learns during training.

\noindent It is important to note that while some existing methods may outperform ours, \textit{e.g.,} on the US8K dataset due to the use of larger datasets. We argue that these methods often overlook the importance of causal reasoning, tending to improve accuracy rather than introducing counterfactuals as an insightful and novel method in bridging counterfactuals to the audio community. In contrast, our method considers both the data and its underlying causal factors, providing a more robust and insightful audio-text representation by borrowing identification and intervention from the science of causality.

\section{Conclusion and Future Direction}

For the first time, we incorporate a counterfactual framework into the audio domain. We leverage LLMs for counterfactual reasoning by prompt-based intervention on the identified acoustic objects. This integration aims to identify variations in audio-text representations by focusing on natural language as a surrogate for the lack of alternative audio-text data when acoustic waves' origin and root cause vary. Our method exploits human-generated reference captions for surrogate counterfactuals and adopts them to the audio-text pretraining in a triplet model with factual consistency. Counterfactual natural language effectively compensates for the scarcity of comprehensive counterfactual audio data for ethical or feasibility reasons and enhances the distinguishability of audio-text models. Empirical evaluations using the various datasets substantiated the effectiveness of our method. In particular, our method yields a 43\% improvement in top-1 accuracy for open-text tasks. Future research may explore the efficacy of these counterfactuals in challenging existing factual representations and their subsequent impact on audio-text correlation. Another avenue for future work could involve examining various levels of counterfactual reasoning.

\end{spacing}
\newpage
\begin{spacing}{0.9} 
\bibliographystyle{IEEEbib}
\bibliography{refs}
\end{spacing} 

\end{document}